# AGGLOMERATION AND INTERREGIONAL MOBILITY OF LABOR IN PORTUGAL

**Vitor João Pereira Domingues Martinho**

Unidade de I&D do Instituto Politécnico de Viseu
Av. Cor. José Maria Vale de Andrade
Campus Politécnico
3504 - 510 Viseu
**(PORTUGAL)**
e-mail: vdmartinho@esav.ipv.pt

**ABSTRACT**

The aim of this paper is to analyze the relationship between inter-industry, intra-industry and inter-regional clustering and demand for labor by companies in Portugal. Is expected at the outset that there is more demand for work where the agglomeration is greater. It should be noted, as a summary conclusion, the results are consistent with the theoretical developments of the New Economic Geography, namely the demand for labor is greater where firms are better able to cluster that is where transport costs are lower and where there is a strong links "backward and forward" and strong economies of agglomeration.

**Keywords:** agglomeration; Portuguese regions; labor; interregional mobility.

## 1. INTRODUCTION

Related to the agglomeration process much has been writing and in varied points of view, that is, in some cases on their causes and others about the process itself (1-2)(Martinho, 2003 and 2011). Works like those of (3)Florence (1948), (4)Perloff et al. (1960), (5)Fuchs (1962), (6)Krugman (1991) and (7)Dumais et al. (1997), for example, have sought to investigate the phenomenon of geographical concentration of economic activity, based on the costs of transport and communication.

In this work, however, we will focus on the relationship between agglomeration and regional demand for labor in industry, analyzing the effect of three forces of agglomeration, transport costs that encourage companies to put up with their activities in regions with relative low cost access to foreign markets; linkage "backward and forward" that give companies an incentive to put near their buyers and suppliers; and economies of agglomeration effects ("spillover") that tend to reinforce the concentration of economic activity, ie, companies benefit from being together, even if they have business relationships with each other and can benefit from, for example, experience accumulated by other companies. It is considered therefore that these three forces create the conditions for which there is agglomeration, through processes of growth pattern circular and cumulative, and that where there is overcrowding there is greater demand for labor by firms (8)(Hanson, 1998).

For transport costs, Krugman (1991), for example, showed that the interaction between economies of scale at the firm level and transportation costs can explain the formation of cities. (9)Krugman and Livas (1992), on the other hand, showed that the size and location of cities is conditioned by the degree of openness of the entire economy. Finally, (10)Rauch (1993) found that transportation costs determine the volume of trade within the region and between countries. Transport costs help to inquire about the inter-regional agglomeration, since one of the purposes of the agglomeration of population and economic activity in a given region is the cost savings of transport and communication.

For the linkages "backward and forward" considered in this work, to stress the fact that (11)Venables (1996) and (12)Krugman and Venables (1995), have formalized the concepts of (13)Hirschman (1958), where the vertical relationships between industries create a model of interdependence the location of industry. Following this expansion of an industry contributes to the expansion of other industries in the same place, which are connected with this commercially, either upstream or downstream, i.e., companies who trade with each other intermediate goods benefit from the proximity factor, a time, save on transportation costs and communication. So this is a more simplistic way of looking this up, i.e. only between industries, because in much of the works related to economic geography the links "backward and forward", representing the centripetal forces, are considered in a more global way and not only between commercially related industries upstream and downstream. With these connections "backward and forward" is intended to analyze the inter-industry clustering.

To reinforce what has been stated previously, noted also that recent theories about the location of economic activity have three elements in common, increasing returns to scale, transportation costs and communication costs and congestion. The new aspect of the industry cluster in this work relates to the fact that companies share links "backward and forward" to each other in the exchange of intermediate goods, as previously mentioned, the centripetal forces, and not only with employees who are also consumers, as usually stated in most of the work on these issues.

## 2. THE ANALYTICAL MODEL OF THE RELATIONSHIP BETWEEN AGGLOMERATION AND REGIONAL DEMAND FOR LABOR

To analyze the relationship between inter-industry clustering, intra-industry and inter-regional and regional demand for labor, we tried to develop an equation that relates the demand relative for labor regional by the companies (represented by employment in each manufacturing Portuguese in a given region of the country, of overall employment in this industry in Portugal) with the three explanatory factors mentioned above the agglomeration, or transportation costs and communication, links "backward and forward" and agglomeration economies, following the procedures of Hanson (1998).

Thus, we obtained the following equation, where we assume that there are positive transportation costs which take the form "iceberg" of Samuelson, i.e. each unit of output of region i sent to another, only a fraction reaches the T destination. The Lij is employment in region i in industry j, and Rem is the salary.

$$\Delta \ln\left(\frac{L_{ijt}}{L_{jt}}\right) = \phi_0 + \phi_1 \ln\left(\frac{\operatorname{Re} m_{ijt-1} / L_{ijt-1}}{\operatorname{Re} m_{jt-1} / L_{jt-1}}\right) + \phi_2 \ln\left(\frac{T_{ij}}{\sum_i \omega_{ijt} T_{ij}}\right) + \phi_3 \ln\left(\frac{L_{ikt-1} / L_{ijt-1}}{L_{kt-1} / L_{jt-1}}\right) + \phi_4 \ln\left(\frac{L_{ijt-1} / L_{it-1}}{L_{jt-1} / L_{t-1}}\right) +$$

$$+ \phi_5 \ln\left(\frac{\sum_{h \neq j} (L_{iht-1} / L_{it-1})^2}{\sum_{h \neq j} (L_{ht-1} / L_{t-1})^2}\right) + \varepsilon_{ijt} - \bar{\varepsilon}_{jt}$$

equation of the relative growth of employment.

Specified in this equation is the relative growth of employment, i.e. employment growth at a Portuguese manufacturing industry in a region of overall employment in this industry in Portugal, which represents the relative regional labor demand by the Portuguese industrial firms. This relative demand for labor is a function of initial conditions in each regional industry in relation to the national industry, where i is the index of the region, j of each manufacturing, operating in Portugal, considered in the sample used, k the total manufacturing and h the total manufacturing except j. Considered to be the explanatory variables lagged one period to analyze the influence of each factor present in a given period in search of work the following year and to avoid introducing bias due to simultaneity.

## 3. STATISTICAL DATA USED

Taking into account the variables of the model presented previously, we used temporal statistical data of the five regions of mainland Portugal, from the regional database of Eurostat statistics (Eurostat Regio of Statistics 2000), relating to salaried employees, regional and national in the manufacturing sector and overall economic activity, the nominal wages in manufacturing and the flow of goods from each of the regions of mainland Portugal to Lisboa e Vale do Tejo. It is this region from the outset as a potential place of agglomeration, given its characteristics, namely, the fact that the highest salaries, to be regions where there are more employees in manufacturing and the region to be associated with higher flows of goods. Nominal wages are solely those of the manufacturing industry, given the emphasis that is given to the sector of manufactured goods, since it is the sectors that produce tradable goods mostly. The regional flow of goods intended to be a "proxy" to transport costs, given that this is an indirect way to measure them, as the authors admit this theoretical approach. In the face of it was considered disaggregated data for nine manufacturing industries operating in Portugal and a series from 1986 to 1994, we had a total of 405 observations in the panel.

## 4. THE ESTIMATES MADE

In all estimations made possible with panel data, the best results, according to the theory, are obtained in the estimations with variables "dummies", with differences and random effects estimates are given in the table below. Were considered 45 "dummies", one for each individual, since the data relate to five regions and nine manufacturing industries. In Table 1 each line refers to an industry for the five regions considered, and each column refers to a region for the nine industries, always in the order mentioned above. Thus, the sixth line of "dummies" refers to food that is not used for worsen statistically the estimation results, possibly because it was an industry with specific features, since the amount depends on agriculture.

However, the Hausman test indicates that the best results are the estimates of fixed effects. Anyway, we present the results obtained in the three estimations, even to serve as a comparison. Explore will be, above all, the results obtained in the estimations with variables "dummies" because they were statistically more satisfactory. It should be noted also that the dependent variable was not considered in growth rate, since this way the results were statistically weaker. Were also carried out simulations considering the variables in the equation productivity and unemployment, but worsen the results statistically and induce the appearance of strange signs for the coefficients, we chose not to consider.

**Table 1:** Estimation of the equation for employment

| | LSDV[1] | | | | | D[2] | GLS[3] |
|---|---|---|---|---|---|---|---|
| | | | | | | | $\phi_0$ -1.878[5]* (-4.424)[6] |
| | $D_1$[4] -0.970[5] (-0.379)[6] | $D_2$[4] -1.575[5] (-0.616)[6] | $D_3$[4] -1.470[5]* (-0.575)[6] | $D_4$[4] 6.782[5] (2.579)[6] | $D_5$[4] (b) | | |
| | $D_6$[4] -0.620[5] (-0.242)[6] | $D_7$[4] -0.576[5] (-0.225)[6] | $D_8$[4] -1.910[5] (-0.748)[6] | $D_9$[4] -8.344[5]* (-3.229)[6] | $D_{10}$[4] -4.204[5] (-1.634)[6] | | |
| | $D_{11}$[4] 1.967[5] (0.768)[6] | $D_{12}$[4] -3.757[5] (-1.465)[6] | $D_{13}$[4] -1.344[5] (-0.526)[6] | $D_{14}$[4] -0.100[5] (-0.039)[6] | $D_{15}$[4] -10.038[5]* (-2.553)[6] | | |
| | $D_{16}$[4] 0.658[5] (0.257)[6] | $D_{17}$[4] 0.792[5] (0.310)[6] | $D_{18}$[4] -3.394[5] (-1.327)[6] | $D_{19}$[4] -4.587[5]** (-1.763)[6] | $D_{20}$[4] -2.234[5] (-0.868)[6] | | |
| | $D_{21}$[4] 0.986[5] (0.385)[6] | $D_{22}$[4] -0.780[5] (-0.305)[6] | $D_{23}$[4] -2.052[5] (-0.803)[6] | $D_{24}$[4] 0.149[5] (0.058)[6] | $D_{25}$[4] -1.317[5] (-0.512)[6] | | |
| | $D_{26}$[4] (c) | $D_{27}$[4] (c) | $D_{28}$[4] (c) | $D_{29}$[4] (c) | $D_{30}$[4] (c) | | |
| | $D_{31}$[4] 0.223[5] (0.087)[6] | $D_{32}$[4] -2.587[5] (-1.009)[6] | $D_{33}$[4] -3.329[5] (-1.303)[6] | $D_{34}$[4] -8.367[5]* (-3.240)[6] | $D_{35}$[4] -13.384[5]* (-3.314)[6] | | |
| | $D_{36}$[4] -0.073[5] (-0.029)[6] | $D_{37}$[4] -0.685[5] (-0.268)[6] | $D_{38}$[4] -1.420[5] (-0.556)[6] | $D_{39}$[4] -5.311[5]* (-2.049)[6] | $D_{40}$[4] -1.921[5] (-0.747)[6] | | |
| | $D_{41}$[4] 0.881[5] (0.344)[6] | $D_{42}$[4] -1.352[5] (-0.529)[6] | $D_{43}$[4] -3.327[5] (-1.302)[6] | $D_{44}$[4] -6.699[5]* (-2.591)[6] | $D_{45}$[4] -7.784[5]* (-3.025)[6] | | |
| | $\phi_1$ 0.119[5]* (2.086)[6] | | | | | $\phi_1$ 0.122[5]* (2.212)[6] | $\phi_1$ 0.112[5]* (1.973)[6] |
| | $\phi_2$ 0.018[5]** (1.879)[6] | | | | | $\phi_2$ 0.012[5] (1.286)[6] | $\phi_2$ 0.022[5]* (2.179)[6] |
| | $\phi_3$ 1.301[5]* (3.241)[6] | | | | | $\phi_3$ 1.127[5]* (2.918)[6] | $\phi_3$ 0.979[5]* (2.443)[6] |
| | $\phi_4$ 0.731[5]* (1.989)[6] | | | | | $\phi_4$ 0.661[5]** (1.867)[6] | $\phi_4$ 0.549[5] (1.492)[6] |
| | $\phi_5$ -0.759[5]* (-4.357)[6] | | | | | $\phi_5$ -0.744[5]* (-4.525)[6] | $\phi_5$ -0.581[5]* (-3.401)[6] |
| $R^2$ adjusted | 0.987 | | | | | 0.217 | 0.683 |
| Durbin-Watson | 2.298 | | | | | 2.086 | 2.068 |
| Hausman Test Chi-square | 7777.548*(a) | | | | | | |

**(1) Estimation with 45 variables "dummies", one for each manufacturing industry, (2) Estimation with differences, (3) Estimation with random effects, (4) Variables "Dummies" (5) value of the coefficient, (6) T - statistic * coefficient statistically significant at the 5% level, ** coefficient statistically significant at 10% (a) reject the hypothesis of random effects, (b) not considered this "dummy" values to present strangers: (c) Do not consider these "dummies" by statistically worse results**

    For variable coefficients "dummies" to mention that vary in terms of value and significance, particularly between regions, in particular, for the last two regions that are the Alentejo and the Algarve. This means that these two regions have significant differences in economic structures relatively to the other three, namely, that the Alentejo, because this region has little economic activity and the Algarve have, especially tourism.

    On the other hand, the equation for employment (employment in each Portuguese manufacturing industry in a given region of the country for employment in this industry in Portugal) gives satisfactory results in terms of statistical significance of the coefficients of the "no dummies," the degree of accuracy of adjustment and

autocorrelation of errors. To highlight the fact that almost all coefficients of the variables "no dummy" with an elasticity less than unity, with the exception of links to the "backward and forward", which indicates the importance of these linkages in explaining the relative employment. Analyzing the results of the estimation for the variables "no dummy" there is, as might be expected, given the developments of the New Economic Geography, there is a positive relationship between relative employment and relative nominal wages (salaries regional industrial concerning the national industrial wages), the positive effect is confirmed also in relation to the links "backward and forward" (ratio between the number of employees in total manufacturing in each region and the number of employees in each manufacturing considered in this region for the same ratio at national level) and in relation to economies of agglomeration (ratio between the number of employees in each manufacturing industry in a given region and the total number of employees in all economy of that region, on the same ratio found at national level). On the other hand, confirms the negative relationship between transport costs and demand on employment (assuming that costs of transport and cargo flows vary inversely) and the ratio of the distribution of employment across industries (the ratio of the sum square of the number of employees in total manufacturing (other than that is being analyzed) of a given region and the total number of employees throughout the economy of this region, for the same sum considered at national level). Note, finally, that all coefficients are statistically significant to 5% and 10% and the high value of the degree of accuracy of adjustment.

## 5. CONCLUSIONS

Given the statistical analysis performed, it appears that for the regions, we can safely say is that on the one hand, the Norte specializes practically in the textile industry and, second, Lisboa e Vale do Tejo has a wide variety of industries, but has the best values for the explanatory factors of regional demand for labor in industry. Complementing the analysis of the data and the estimation results, noted also that transportation costs are important because, in addition to the statistical significance of the coefficient associated with this variable, the regions closer to the Lisboa e Vale do Tejo has the greater flow of goods to this region. There are also links "backward and forward" between the different manufacturing industries, there are economies of agglomeration effects ("spillover") between firms and industries and there is a distribution, more or less uniform, of the employment across different industries. Productivity and unemployment have no influence in explaining the demand for industrial work at the regional level, given the results obtained in the estimations, when these variables were considered.

Following this, it appears that the results are consistent with the developments of the New Economic Geography which emphasizes the transportation costs, which are based on a set of other explanatory variables, such as links "backward and forward," i.e., buyers and suppliers are looking to be together in order to save on transport costs and agglomeration economies. Note that Alfred Marshall in 1920 had made reference to such connections, when the modeling of increasing returns that explain the spatial concentration, has proposed a threefold classification that is presented below. In modern terminology, he advocated that the locations appear on the face of industrial purposes "spillovers," the advantages of specialized markets and links "backward" and "forward" associated with large local markets. Although all these three forces are clearly operating in the real world, the New Economic Geography has generally ignored the first two, mainly because they are difficult to model in an explicit manner.

Finally, in light of the foregoing, it is noted that Lisboa e Vale do Tejo is a local potential agglomeration of population and economic activity (as we suppose), since the flow of goods is higher in regions closer to this region , relative employment is higher in the previous period where wages were higher (knowing that it is in Lisboa e Vale do Tejo that wages are higher) and to characterize inter-industry and inter-industry described.

There are other factors which, although implicit, would be worthwhile to disaggregate further analysis, as the entrepreneurial dynamics ((14)Martinho, 2010a and (15)Martinho, 2010b) and the spatial effects ((16)Martinho, 2011a, and (17)Martinho, 2011b).